\documentclass[11pt]{article}
\usepackage[utf8]{inputenc}
\usepackage{geometry}
\usepackage{amsmath, amssymb}
\usepackage{booktabs}
\usepackage{verbatim}
\usepackage{xcolor}
\usepackage{caption}
\usepackage{graphicx}
\usepackage{enumitem}
\usepackage{amsmath}
\usepackage{amssymb}
\usepackage{bm}
\usepackage{xurl}
\usepackage{amsthm}
\usepackage{mathtools} 
\usepackage[most]{tcolorbox} 
\usepackage{natbib}
\newtcolorbox{promptbox}[1]{
  colback=gray!5!white,
  colframe=gray!50!black,
  title=\textbf{#1},
  fonttitle=\bfseries,
  boxrule=0.5pt,
  arc=2pt
}
\newtcolorbox{proofbox}[1]{
  colback=blue!3!white,
  colframe=blue!40!black,
  title=\textbf{#1},
  fonttitle=\bfseries,
  boxrule=0.5pt,
  arc=2pt,
  breakable,
  fontupper=\small
}
\usepackage{hyperref}
\usepackage{float}

\usepackage[most]{tcolorbox}
\usepackage[nameinlink,noabbrev]{cleveref}

\tcbuselibrary{theorems,breakable}

\newtcbtheorem[auto counter]{problembox}{Problem}%
{enhanced,
 breakable,
 colback=gray!10,
 colframe=gray!50!black,
 coltitle=white,
 fonttitle=\bfseries,
 boxrule=0.8pt,
 arc=2pt,
 left=6pt,
 right=6pt,
 top=6pt,
 bottom=6pt,
 title filled=true,
 colbacktitle=gray!30!black}{prob}

\newtcbtheorem[auto counter]{contextbox}{Context}%
{enhanced,
 breakable,
 colback=gray!10,
 colframe=gray!50!black,
 coltitle=white,
 fonttitle=\bfseries,
 boxrule=0.8pt,
 arc=2pt,
 left=6pt,
 right=6pt,
 top=6pt,
 bottom=6pt,
 title filled=true,
 colbacktitle=gray!30!black}{ctx}

\crefname{problembox}{Problem}{Problems}
\Crefname{problembox}{Problem}{Problems}
\crefname{contextbox}{Context}{Contexts}
\Crefname{contextbox}{Context}{Contexts}

\newcommand{\y}{\bm{y}}
\newcommand{\X}{\bm{X}}

\newcommand{\Z}{\bm{Z}}

\newcommand{\bmeps}{\bm{\epsilon}}

\newcommand{\bmalpha}{\bm{\alpha}}

\newcommand{\tr}{\mbox{tr}}
\newcommand{\bbR}{\mathbb{R}}

\newcommand{\bmSigma}{\bm{\Sigma}}

\ifx\theorem\undefined
\newtheorem{theorem}{Theorem}
\fi
\newcommand{\Var}{\operatorname{Var}}

\title{Statistical Proof as a Window into Human-AI Collaboration: Practical Insights and a Community Agenda}

\author{Xiaojing Sun\thanks{Purdue University} \and Huayu Tang\footnotemark[1] \and Buxin Su\thanks{University of Pennsylvania} \and Mateo Matijasevick\footnotemark[1] \and Chong Wu\thanks{University of Texas MD Anderson Cancer Center} \and Fei Xue\footnotemark[1]~\thanks{These authors jointly supervised this work. Correspondence: \texttt{feixue@purdue.edu;bxzhao@wharton.upenn.edu}} \and Bingxin Zhao\footnotemark[2]~\footnotemark[4]}

\begin{document}

\maketitle

\begin{abstract}

Large language models (LLMs) are increasingly woven into expert cognitive work in daily research, yet we know little about how human expertise should adapt when an AI system can execute substantial technical reasoning on its own. Here we use statistical proof development, a demanding and structured form of expert reasoning, as a window into this broader question. Drawing on day-to-day proof problems, we find that current general-purpose LLMs occupy a useful but limited role: they can execute technical components given a precisely formulated problem and targeted guidance, but become unreliable when the problem is open-ended or requires a long reasoning chain with multiple interdependent steps. This execution-strategy gap is rooted in what makes research-level statistical proof distinctive: unlike pure mathematics, where problems arrive pre-formulated and solutions often demand entirely novel techniques, statistical proof requires first modeling a scientific question into a statistical framework with appropriate assumptions, and then identifying and adapting the right strategy from a repertoire of reusable domain-specific tools. Each step involves judgments that require deep expertise in both the statistical literature and the real-world context being modeled. For human-AI collaboration in such cognitive work, current AI assistance does not reduce the need for human expertise; it relocates that expertise to where human decision-making matters most, such as problem formulation and verification of AI-generated results, and may raise the bar for both, since human users must now make rapid decisions alongside AI tools. These findings yield practical suggestions for how statisticians can structure AI-assisted proof workflows in their daily research, and point to a broader community agenda for shared resources, better AI tools, and training the next generation of researchers. Using statistical proof as a window, our study carries implications for how experts structure human-AI collaboration in technical cognitive domains more broadly.

\end{abstract}

\section{Introduction}

Proof is the backbone of statistical research, where theory and real-world scientific problems are deeply intertwined. Through rigorous mathematical argument, statisticians justify new estimators and procedures used in data analysis, clarify the assumptions under which they are valid, and explain how performance depends on model structure, dataset features, and noise \citep{zhao2025statistical}. These arguments are often directly motivated by interdisciplinary applications such as in genomics, economics, and medicine, where the choice of assumptions needed to form a statistical framework and the interpretation of results must remain grounded in the underlying science.
Together, these features make statistical proof development a demanding form of expert cognition: it requires scientific judgment about how to formulate a real-world question as a tractable statistical problem, domain expertise to navigate a broad landscape of proof strategies, and the experience to know which strategy fits a new problem and context, capabilities that can take years to acquire even for experienced researchers.
For graduate students, acquiring this expertise is also challenging: they must learn to formulate real-world questions as precise statistical problems, identify appropriate proof strategies, and execute rigorous arguments, skills that are currently transmitted slowly and unevenly through textbooks, previous literature, and one-on-one guidance from their advisors. These bottlenecks slow both the development of new statistical methodology and the transmission of proof expertise to the next generation of statisticians.

Large language models (LLMs) are now reshaping how scientists conduct research across many disciplines. 
Notably, LLM-based AI systems are increasingly capable of mathematical reasoning and proof assistance \citep{wang2026survey,ju2026ai,ren2025deepseek, RomeraParedes2024FunSearch}, and recent progress in Olympiad-style problem solving, formal verification, and research-level reasoning suggests that these systems may become useful tools for mathematical research more broadly \citep{Trinh2024AlphaGeometry,Hubert2026AlphaProof,Woodruff2026GeminiCaseStudies,Feng2026Aletheia}. 
The statistics community is no exception: researchers are increasingly asking whether AI can assist with proof development, the central intellectual task of statistical research \citep{Abouzaid2026FirstProof}. From a practical standpoint, a key question is whether current general-purpose LLMs, which are now widely accessible to everyday researchers, can provide meaningful assistance with proof tasks in day-to-day statistical research. Progress on this question would have direct implications not only for how statisticians work, but also for how graduate students are trained, how proof expertise is acquired and shared, and how the community allocates its intellectual effort. Yet despite the growing presence of LLMs in statistical research workflows, their boundaries as proof assistants remain poorly understood: which tasks they can meaningfully assist with, and which remain out of reach?

These open questions in statistical proof provide a useful lens for a broader question, now facing many areas of expert work, about human-AI collaboration: as AI systems take on substantial technical reasoning in daily research, how should human expertise adapt? This question reaches well beyond statistics, into other fields where experts are beginning to delegate technical execution to AI while remaining responsible for the direction and outcome. Statistical proof is a demanding and highly structured form of expert cognition, which makes it a clear setting in which to study this question: it allows us to ask not just whether AI can help, but specifically where in the process AI assistance is reliable and where human judgment remains essential. The practical insights that emerge are not a narrow statistical finding; they speak to many expert domain in which technical work is increasingly being reshaped by human-AI collaboration.

Although mathematics is a related domain where AI proof assistance has been increasingly tested, the insights and progress in AI for mathematical reasoning do not transfer automatically to statistical proof. Statistical proof tasks differ from most mathematical proof tasks in at least two important respects that determine when and how AI assistance can be effective. First, statistical problems are inherently closer to real-world applications than most mathematical problems \citep{zhao2025statistical}. Assumptions in statistical proofs are rarely mathematically simple or clean; they must be chosen to be both tractable and interpretable, and consistent with the scientific problems being modeled, a judgment that has no direct analogue in pure mathematics, where problems typically arrive pre-formulated. Second, statistical proofs less commonly demand entirely novel proof approaches; instead, they predominantly build on a repertoire of reusable domain-specific strategies that need to be identified and adapted to each new problem and context. Effective AI assistance for statistical proof thus requires a different set of capabilities, priorities, and resources from those needed for mathematical reasoning \citep{SeedProver2025, ProverAgent2025}.

\begin{figure}[t]
    \centering
    \includegraphics[width=\textwidth]{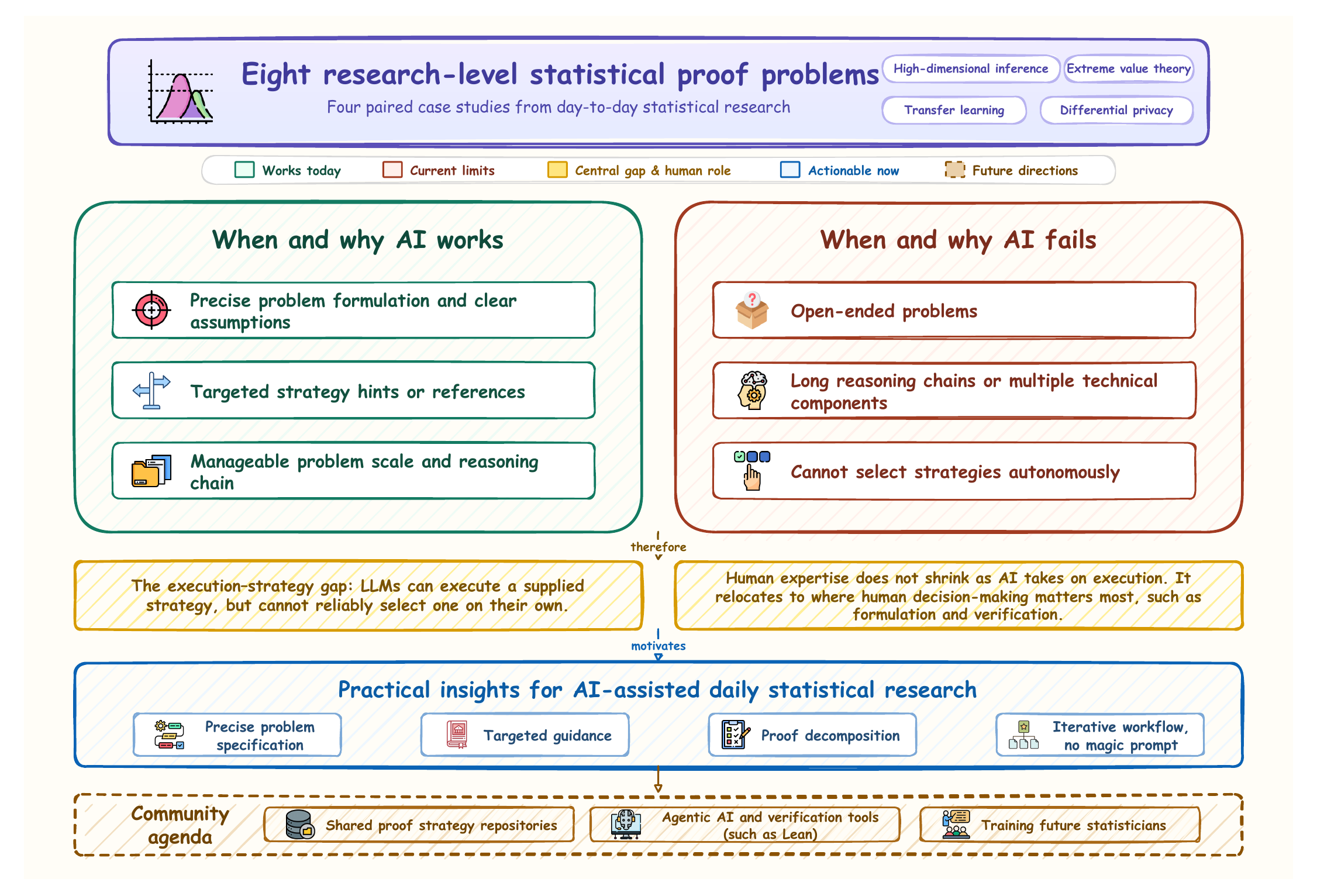}
   % \caption{Overview of AI-assisted statistical proof writing.}
   \caption{Study overview. Four paired case studies covering eight research-level proof problems from day-to-day statistical research reveal a consistent pattern: current general-purpose LLMs can execute a proof strategy when one is supplied but cannot reliably select one on their own. This execution-strategy gap (yellow) emerges from examining when and why AI assistance works (green, Section~\ref{work}) and when and why it fails (red, Section~\ref{notwork}). The gap carries a parallel implication for the human role (yellow): AI assistance does not reduce the expertise that statistical proof demands but relocates it to problem formulation and verification of AI-generated results, where the bar on human judgment may rise. Together, these findings provide practical insights for individual statisticians working with these tools today (blue) and motivate a broader community agenda for future AI developments (amber, dashed).}
    \label{fig:workflow}
\end{figure}

In this paper, we investigate when and how current general-purpose LLMs succeed and fail as proof assistants in day-to-day statistical research. 
Through paired case studies drawn from active research problems across several areas of modern statistical research, we identify a consistent pattern: current general-purpose LLMs can assist with substantial components of statistical proof when the problem is precisely stated and supplemented with targeted guidance, but become unreliable when the problem is open-ended, requires a long reasoning chain, or demands simultaneous control of multiple technical components, settings in which the model must identify and adapt an appropriate strategy on its own. This execution-strategy gap is currently a primary determinant of whether AI assistance succeeds or fails on research-level statistical proof tasks. These observations indicate that the demands on human expertise do not shrink as AI takes on more of the execution; rather, they relocate to the points of greatest judgment, such as problem formulation and verification of AI-generated results, and may grow more exacting as experts must make rapid decisions alongside AI tools. These demands may also grow more exacting, as experts must now make rapid judgments alongside AI tools. Based on these observations, we discuss practical lessons for individual statisticians, covering problem formulation, proof decomposition, rapid verification, and the use of targeted guidance, and outline a broader community agenda: building shared repositories of reusable proof strategies, developing agentic AI and proof assistants, and training future statisticians to interact effectively with AI tools. As AI systems take on more of the technical execution across many expert domains, our practical insights from statistical proof provide an early and clear picture of what this shift demands of human expertise.

\section{Study design and key findings}
\label{sec:overview}

We evaluate AI assistance for statistical proof development through four paired case studies that together cover eight research-level proof problems. Within each pair, one problem illustrates a setting in which current general-purpose LLMs provide useful proof assistance, while the other illustrates a setting in which they become unreliable. Both problems are drawn from the same broader research context, enabling direct comparison and practical insights into when and how these tools can be used most effectively in statistical research. 
Specifically, the four pairs cover high-dimensional inference, extreme value theory, transfer learning, and differential privacy, together reflecting the breadth of modern statistical research \citep{XueZhao2025LDSC,Wainwright2019,LedfordTawn1996,deHaan2006,Nelsen2006}. 
Most proof problems arise from ongoing research projects without publicly available proofs. The only exception is the high-dimensional inference pair, where a related paper  \cite{XueZhao2025LDSC} is publicly available, yet the AI models arrived at correct results through proof strategies distinct from those in that paper, suggesting the model followed a genuinely different proof route rather than reproducing it.
An overview of the study design is shown in Figure~\ref{fig:workflow}.

For each problem, we evaluated three widely used general-purpose LLMs: GPT-5.4 Thinking, Gemini 3.1 Pro, and Claude Opus 4.6. The qualitative conclusions were remarkably consistent across all three models, suggesting that the patterns we identify reflect systematic properties of current general-purpose LLMs. 
Unless stated otherwise, we focus our illustrations on the results of GPT-5.4 Thinking, with model-specific outputs for all three models reported in the supplementary material.
Our goal is not to rank or compare these models, but to extract practical insights for statisticians working with these tools today, and to outline a broader community agenda for advancing AI-assisted statistical proof development. 

Our case studies yield the following practical insights into the capabilities and limitations of current general-purpose LLMs for statistical proof development.
\begin{itemize}
	\item Clarity and precision enable success. Current general-purpose LLMs can assist with substantial components of research-level statistical proof when the problem is stated clearly, the goal is explicit, the reasoning scope is kept at an appropriate scale, and the prompt is written in a precise and professional way. 
	\item  Zero-shot prompting is often insufficient.  Models frequently struggle when the problem depends on newly introduced context \citep{Dou2026CLBench}, requires careful control of modeling assumptions, or demands selection of an appropriate proof strategy. In these cases, models often produce plausible-looking but ultimately incorrect or incomplete derivations.
	\item Modest guidance can make a substantial difference. Concise proof-strategy hints or literature-based methodological references often enable the model to produce a correct and rigorous argument where zero-shot prompting had failed. The right guidance does not need to be extensive, a few targeted sentences are often enough to redirect the model toward a valid proof strategy.
	\item  Execution is more reliable than strategy selection. Current general-purpose LLMs perform considerably better at executing a suitable proof strategy than at identifying one on their own. This asymmetry, between what LLMs can execute and what they can identify, is the primary manifestation of the execution–strategy gap in practice.
	\item  Long reasoning chains and weak guidance lead to failure. AI reliability decreases substantially when the argument requires a long reasoning chain, simultaneous control of multiple technical components, or decision making under weak guidance. In these settings, models tend to introduce convenient but unjustified intermediate assumptions, shift the target of the proof, or produce notation-heavy derivations that obscure rather than illuminate the main argument.
\end{itemize}

Together, these observations suggest that current general-purpose LLMs can be meaningfully integrated into statistical proof development within a structured workflow that specifies the problem clearly and professionally, states assumptions explicitly, decomposes the proof into manageable problems, and introduces targeted guidance when necessary. They also point to a broader need in the statistics community for explicit and accessible resources that organize reusable proof strategies by research area, since shared repositories of proof strategies and literature-based templates could improve both human and AI-assisted proof development. For education, these findings suggest that training future statisticians should include not only mathematical theory but also effective interaction with AI tools: learning to formulate statistical problems precisely, specify context and assumptions, evaluate whether an AI-generated proof uses an appropriate strategy, provide corrective guidance when it does not, and distinguish a plausible-looking but inaccurate derivation from a rigorous one.

Beyond these specific lessons, our case studies point to more general insights about the role of human expertise in human-AI collaboration for statistical proof.

\begin{itemize}
	\item Defining a good proof task is a growing bottleneck. As AI lowers the technical barrier to executing a proof, the difficulty shifts to what AI currently cannot do: knowing what should be proved in a scientific context, why that problem is important, what assumptions are appropriate, and what result would be scientifically meaningful. This requires a chain of human decisions, from selecting a worthwhile research question, to formulating it as a precise statistical problem, to conveying that precision to the AI model in efficient human-AI conversations. Human researchers who can make these decisions fluently extract more value from AI assistance and direct it toward more meaningful research outputs.
	\item  Live human verification makes the workflow productive. In proof development, the realistic workflow is iterative, closer to an engaged human-AI conversation than to a one-shot prompt query. At every step, the human user must rapidly diagnose whether the AI-generated proof is valid. This diagnosis is often not a matter of mathematical correctness alone; it is about whether the proof answers the intended question, whether unjustified assumptions have been quietly introduced, and whether the result is sharp and scientifically useful. A good human expert can often tell quickly whether a generated proof is on the right track, has changed the target, or has produced a result that is mathematically valid but scientifically weak. This kind of rapid proof triage is what makes the iterative workflow actually work: an AI model can only be corrected by guidance if a human already knows what went wrong. Consistent with the first point, AI does not reduce the expertise statistical proof demands; it relocates and intensifies it.
\end{itemize}

The contributions of this paper are therefore fourfold. First, we provide a set of paired research-level case studies that characterize both productive and unreliable scenarios for general-purpose LLMs as proof assistants in statistical research. Second, we identify an execution-strategy gap that helps explain when AI assistance succeeds and fails, and show that this gap reflects a broader pattern: AI does not reduce the human expertise that statistical proof demands, but relocates it to where human decision-making matters most. Third, we extract practical insights for integrating AI tools into daily statistical research workflows. Fourth, we outline a broader community agenda and highlight the opportunities and challenges this shift in human-AI collaboration creates. We view this as an evolving effort and will continue to update our findings as additional proof problems, AI models, and tools become available.

\section{When and how AI assistance succeeds}\label{work}

Despite coming from different areas of statistical research, the four successful cases reveal a consistent pattern: AI assistance works best when the proof task is precisely scoped, the modeling assumptions are made explicit, and the reasoning chain is kept at a manageable scale. In three of the four cases, zero-shot prompting, asking the model to solve the problem directly from the problem statement alone, without any additional domain-specific guidance, was insufficient. In those cases, however, relatively modest and targeted guidance, such as a short strategy hint, a key literature reference, or an important clarifying instruction, was enough to redirect the model toward a correct and rigorous argument. These cases illustrate that the barrier to effective AI assistance in statistical proof is often not mathematical difficulty alone, but the availability of the right proof strategy at the right moment.

\subsection{Variance calculation in high-dimensional inference}\label{positveLDSC}

Our first example comes from linkage disequilibrium score regression (LDSC), a widely used method for estimating trait heritability from genome-wide association study (GWAS) summary statistics \citep{bulik2015ld}. 
The specific task is a technical problem in the high-dimensional asymptotic analysis of LDSC: computing the exact finite-sample variance of the oracle LDSC estimator, where ``oracle" means that the linkage disequilibrium (LD) scores are treated as known. This problem is mathematically nontrivial but sufficiently self-contained, making it a natural task of whether AI can carry out a rigorous derivation when the statistical setting is stated clearly.
The prompt and problem setup are summarized in the following boxes, with full derivations provided in the supplementary material.

\begin{promptbox}{Zero-shot prompt 1}
You are an expert in mathematical statistics and genetics. 
The following is a research-level problem. Please make the best effort to
provide a rigorous and complete formal proof for the problem.

\textbf{Problem:} The problem stated in Problem~\ref{prob:varianceLDSC}

\textbf{Emphasize:} (i) The genetic effects are fixed. Do NOT assume that they are random effects. 
(ii) Please define notations carefully and use only  natural notations.
(iii) Write the output in LaTeX language 
 using the standards of rigor and scholarship that
prevail in the mathematical literature.
\end{promptbox}

\begin{problembox}{}{varianceLDSC}
This is a sub-problem of the high-dimensional inference for the linkage disequilibrium score regression (LDSC) in the genome-wide association studies (GWAS).

Suppose that we have a GWAS cohort with $\X\in \bbR^{n\times p}$ consisting samples of $p$ genetic variants and $\y\in\bbR^{n}$ consisting samples of the outcome.
Let $\bmSigma$ be the covariance matrix of the $p$ genetic variants.
Define the LD score $\ell_{j}=\sum_{i=1}^{p}\bmSigma_{ji}^2$ and the estimated marginal covariance $\widehat{a}_j=n^{-1}\cdot \X_{j}^T\y$, where $\bmSigma_{ji}$ is the $(j,i)$ element of $\bmSigma$ and  $\X_{j}$ is the $j$-th column of $\X$ for $i=1, \dots, n$ and $j=1, \dots, p$.
Let $\widehat{w}=(\widehat{a}_1^2,\dots, \widehat{a}_p^2)^\top$, $\ell=(\ell_{1},\dots,\ell_{p})^\top$, 
and
\[
\mu_{\ell}=\frac{1}{p}\,\bm{1}_p^\top \ell,\qquad
\hat\mu_{w}=\frac{1}{p}\,\bm{1}_p^\top \hat w.
\]
Define the (centered) ``oracle'' LDSC estimator
\[
\tilde\sigma_\alpha^2
= \frac{(\ell-\mu_{\ell}\bm{1}_p)^\top(\hat w-\hat\mu_{w}\bm{1}_p)}{(\ell-\mu_{\ell}\bm{1}_p)^\top(\ell-\mu_{\ell}\bm{1}_p)}.
\]

We assume that
\[
    \y= \X\bmalpha+\bmeps,
\]
where  $\bmalpha=(\alpha_1,\ldots, \alpha_{p})^{T}$ 
are unknown fixed genetic effects and $\bmeps=(\epsilon_{1},\ldots,\epsilon_{n})^T$ consists of mean-zero normal random error terms with variance $\sigma_{\epsilon}^2$.
We also assume that all the genetic variants follow a multivariate normal distribution with a block-diagonal structured covariance matrix.

Calculate the variance of $\tilde\sigma_\alpha^2$.
Write down the final results in terms of $n$, $\bmSigma$, $\ell$, $\bmalpha$, and $\sigma_{\epsilon}^2$.
\end{problembox}

The model produced a complete derivation whose final variance formula agrees with the result we derived independently in \cite{XueZhao2025LDSC}. Notably, despite the public availability of our paper, the AI-generated proof arrived at the correct answer through a different proof strategy, suggesting the result was not a simple retrieval of the published work.

\begin{proofbox}{AI-generated proof 1}
\paragraph{Final variance formula.}
Let
\[
c=\ell-\mu_\ell\mathbf{1}_p,\qquad C=\operatorname{diag}(c),\qquad
\Delta=\sum_{j=1}^p(\ell_j-\mu_\ell)^2,
\]
and
\[
\tau^2=\bmalpha^\top\bmSigma\bmalpha+\sigma_\varepsilon^2.
\]
Then
\begin{align*}
%\boxed{
\Var(\widetilde{\sigma}_{\alpha}^2)
=&
\frac{1}{\Delta^2}
\left[
\frac{8(n+2)(n+3)}{n^3}B^2
+\frac{4(n+2)(n+4)}{n^3}\tau^2R
+\frac{8(n+2)}{n^3}\tau^2BS \right.\\
&\left.+\frac{2(n+2)}{n^3}\tau^4U
+\frac{2}{n^3}\tau^4S^2
\right],
%}
\end{align*}
where
\[
B=\bmalpha^\top\bmSigma C\bmSigma\bmalpha,
\]
\[
S=\tr(C\bmSigma)-\frac{\bmalpha^\top\bmSigma C\bmSigma\bmalpha}{\tau^2},
\]
\[
R=
\bmalpha^\top\bmSigma C\bmSigma C\bmSigma\bmalpha
-\frac{(\bmalpha^\top\bmSigma C\bmSigma\bmalpha)^2}{\tau^2},
\]
\[
U=
\tr\!\bigl((C\bmSigma)^2\bigr)
-\frac{2\,\bmalpha^\top\bmSigma C\bmSigma C\bmSigma\bmalpha}{\tau^2}
+\frac{(\bmalpha^\top\bmSigma C\bmSigma\bmalpha)^2}{\tau^4}.
\]

This is the exact finite-sample variance of the centered oracle LDSC estimator under the Gaussian design model with fixed effects \(\bmalpha\).
\end{proofbox}

This example illustrates that zero-shot prompting can be sufficient for research-level statistical proof when the task is precisely scoped and the prompt is written with care. 
Although zero-shot prompting ultimately succeeded here, the experience of getting there yields several practical lessons, and shows that human expertise plays a strong role during problem formulation. First, the problem must be stated precisely and the relevant assumptions made explicit. When the setup is underspecified, the model tends to introduce additional assumptions that are mathematically convenient but inconsistent with the intended statistical framework and scientific context.
Second, when the intended modeling framework differs from what dominates the relevant literature, the prompt must do more than state the correct assumption. It must also explicitly rule out the incorrect default.
Specifically, the critical assumption here was that the genetic effects are fixed rather than random. Because the LDSC literature is predominantly developed under a random-effects framework \citep{bulik2015ld,jiang2023high}, the model defaulted to that framework even when a fixed-effects setup was provided in the prompt, producing a derivation that appeared superficially reasonable but was answering the wrong question. The fixed-effects assumption is essential here because our goal is an exact finite-sample variance, which requires treating the genetic effects as deterministic rather than stochastic. Explicitly instructing the model to treat the genetic effects as fixed, and to avoid the random-effects framework, was necessary to obtain a correct derivation. This is a subtle but serious pitfall: for researchers less familiar with the distinction between fixed and random effects in this context, an AI-generated proof under the wrong framework may appear valid on the surface while being fundamentally misaligned with the intended problem.

\subsection{Scaling invariance of angular tail dependence in copulas}
\label{positive example:copula}
Our second example comes from an ongoing project on copulas \citep{Sklar1959} and angular tail dependence, which is representative of a broad class of questions in extreme value theory \citep{deHaan2006}. 
The task is to prove that a limit-based measure of lower-tail behavior \citep{joe2014dependence, hua2011tail, su2017general}, designed to quantify dependence along different angular paths toward the tail, remains invariant with respect to a scaling parameter under suitable regularity conditions. 
The main difficulty here is strategic rather than computational. The proof requires choosing an appropriate way to compare limits and identifying the regularity conditions needed to justify each step. This is the kind of judgment that current general-purpose LLMs struggle to exercise without guidance. Yet as we show below, a short and targeted proof-strategy hint is often enough to resolve it.

The problem setup is given in the following box. 
The quantity $\kappa_L(\delta, \gamma; C)$ is designed to capture lower-tail behavior along an angular path indexed by $\delta$, while allowing a scale perturbation indexed by $\gamma$.
Establishing invariance with respect to $\gamma$ is a necessary theoretical step before $\kappa_L$ can be interpreted as a coherent and well-defined measure of angular tail dependence.

\begin{problembox}{}{copula}
For a copula $C$ on $[0,1]^2$ and parameters $\gamma>0$, $\delta\in(0,2)$, define
\[
\kappa_L(\delta,\gamma;C)
=
\lim_{u\downarrow 0}
\frac{\log C\!\bigl(\gamma u^\delta,\gamma^{-1}u^{2-\delta}\bigr)}{\log u},
\]
whenever the limit exists.
The claim to be proved is that 
under suitable regularity conditions, $\kappa_L(\delta,\gamma;C)$ does not depend on the choice of scaling parameter $\gamma$ for all $\delta \in (0,2)$.
\end{problembox}

To help the model engage with this problem correctly, we provided not only the specific mathematical question but also the surrounding scientific context, clarifying the motivation of the problem, its role in the broader research project, and why the claimed invariance property is important. They are summarized in the box below. 

\begin{contextbox}{}{copula}
We propose a new class of copulas that can account for {angular full-range tail dependence}, capturing not only asymmetry but also non-exchangeability.
Our proposed copula, denoted by $C_{\rm PM}$, is defined explicitly {in the supplement.}
To theoretically analyze this new model class, we first need a principled measure of tail behavior that can vary with the {direction} (angle) of approach to the tail.
That is the motivation of how we define $\kappa_L(\delta,\gamma;C)$.
This quantity is a generalized lower-tail dependence coefficient~\citep{hua2011tail}.
Our intuition is that, when the limit exists, the resulting angular tail behavior should primarily be governed by the {direction parameter} $\delta$, and should not depend on the {scaling parameter} $\gamma$.
We therefore seek (i) regularity conditions under which such scaling invariance holds and (ii) a rigorous proof under those conditions.
\end{contextbox}

We first provided the model with the problem context, the formal statement to be proved, and a brief prompt instruction.

\begin{promptbox}{Zero-shot prompt 2}
``You are an expert in statistical methodology and mathematical reasoning. Your task is to produce a full formal proof for a problem in an ongoing research project.

\textbf{Problem:} Prove the claim stated in Problem~\ref{prob:copula}. 
In addition, verify that the proposed copula $C_{\rm PM}$ described in Context~\ref{ctx:copula} also satisfies this claim.

\textbf{Context:} The background provided in Context~\ref{ctx:copula}.

Given the problem and context above, give a full formal proof for the problem.
State any assumptions that are needed.
Justify each step and give the complete proof in \LaTeX.''
\end{promptbox}

Without any proof-strategy guidance, the model produced a lengthy and seemingly well-structured derivation. However, it adopted a proof technique rarely used for problems of this type and did not fully engage with the newly introduced research context, leading to several non-rigorous steps, most notably, failing to state the regularity conditions necessary to justify key arguments. This zero-shot attempt therefore did not constitute a valid proof. Notably, this failure was not algebraic: the model made no obvious calculation errors, but did not recognize which proof approach was appropriate for this statistical problem. The main limitation was thus strategic rather than computational.

The standard approach for establishing this type of invariance result is to bound the difference between the two quantities of interest by decomposing it using the triangle inequality and then invoking suitable regularity conditions to control each term \citep{resnick2007heavy}, an approach that statisticians working in extreme value theory would recognize as natural and well-suited to this problem. The model did not adopt this strategy, instead pursuing a less natural and ultimately non-rigorous route. This observation motivated us to supplement the original prompt with two concise proof-strategy hints pointing toward the standard approach, as shown in the box below.

\begin{promptbox}{Guidance prompt 1}
	
	\textbf{Guidance:} 
	To prove that $\kappa_L(\delta,\gamma;C)$ does not depend on $\gamma$, it is enough to show that, for any $\gamma,\tilde{\gamma}>0$,
	\(
	\kappa_L(\delta,\gamma;C)=\kappa_L(\delta,\tilde{\gamma};C).
	\)
	That is, we need to prove the equality of the two limits corresponding to $\gamma$ and $\tilde{\gamma}$.
	
	(i)~To prove equality of these two limits, consider to show that there exists $u_0\in(0,1)$ such that for all $0<u<u_0$, the absolute value of the difference of the corresponding pre-limit quantities is less than or equal to $\varepsilon$ for every $\varepsilon>0$;
	
	(ii)~To use the triangle inequality by adding and subtracting terms inside the absolute value, bounding each term, and stating essential assumptions of the limits (such as type of the convergence) based on our research context and problem.
	
	\textbf{User:} ``Give a {full formal proof}. Justify each step and state any assumptions you use''
	
\end{promptbox}

With these two strategy hints, the model produced a correct and rigorous proof, decomposing the difference into three terms and bounding each by $\varepsilon/3$, with appropriate regularity conditions stated explicitly. 
The same model that had produced a non-rigorous derivation under zero-shot prompting delivered a correct and complete proof once the right strategy was indicated, demonstrating that the bottleneck was not mathematical ability but strategy identification.

This example illustrates a clear instance of the execution–strategy gap: the bottleneck was not mathematical execution but strategy identification. A key reason the zero-shot attempt failed is that the model did not genuinely understand the research context. It processed the problem statement and background but could not situate the proof task within the broader landscape of extreme value theory, and therefore had no principled basis for selecting an appropriate proof strategy. 
This is where human expertise becomes essential. A statistician familiar with the domain can recognize immediately which proof approach is natural for a problem of this type, and supplying that insight in a few targeted sentences can change the entire outcome. Conversely, without such guidance, the model may pursue a plausible-looking but ultimately invalid route, producing a derivation that appears well-structured on the surface but fails to meet the standards of statistical rigor. 
This points to a practical workflow for statisticians: rather than asking the model to develop a proof from scratch, it may be more effective to first identify the appropriate proof strategy, drawing on domain expertise and familiarity with the statistical literature, and then provide that strategy as targeted guidance before asking the model to execute. 

\subsection{Bounding the prediction risk in transfer learning}\label{positive example:TLRisk}

Our third example comes from an ongoing project on transfer learning under a high-dimensional linear regression framework \citep{gu2025robust,li2022transfer}. The goal is to derive an asymptotic upper bound for the prediction risk of a transfer learning estimator that leverages information from a source site to improve prediction at a target site, while accounting for cross-site heterogeneity and feature mismatch. Deriving a sharp and informative bound, rather than a valid but loose one, is the central challenge, as an overly conservative bound provides little insight into how the estimator's performance depends on the key problem parameters. 
The specific context and setup of this problem are described in the boxes below.

\begin{contextbox}{}{tl}
We consider a transfer learning setting consisting of a target site $t$ and a source site $s$.
For each site $m\in\{t,s\}$, the response $y^{(m)}\in\mathbb{R}$ and the covariate vector $x\in\mathbb{R}^p$ satisfy the linear model
\[
y^{(m)} = x^\top \beta^{(m)} + \varepsilon,
\]
where $\beta^{(m)}\in\mathbb{R}^p$ is a random coefficient vector, $x\sim N(0,\Sigma)$, and $\varepsilon\sim N(0,\sigma^2)$ is independent of $x$. The dimension $p$ increases to infinity as the sample size $n$ increases.
\end{contextbox}

\begin{problembox}{}{tl}
Suppose that we only observe the subvector $x_{\mathcal R(t)}$, indexed by $\mathcal R(t)\subseteq [p]$ in the target site. Let $X^{(t)}_{\mathcal R(t)}\in\mathbb{R}^{n_t\times p_t}$ and $Y^{(t)}\in\mathbb{R}^{n_t}$ denote the observed design matrix and response vector, respectively, where $p_t=|\mathcal R(t)|$ and $n_t$ is the target-site sample size.\\

Given an estimator $\hat{\beta}^{(s)}$ of the source-site coefficient vector $\beta^{(s)}$, we define a transfer learning estimator $\hat\omega_{\mathrm{TL}}^{(t)}$ for the coefficient vector corresponding to the observed target features $x_{\mathcal R(t)}$ by
\[
\hat\omega_{\mathrm{TL}}^{(t)}
:=
\arg\min_{\omega}
\left[
\frac1{n_t}
\|Y^{(t)}-X^{(t)}_{\mathcal R(t)}\omega\|_2^2
+\lambda_{\mathrm{TL}}\|\omega-\hat{\omega}_s^{(t)}\|_2^2
\right],
\]
where $\hat\omega_{s}^{(t)} = \Phi^{(t)}\hat\beta^{(s)}$, and $\lambda_{\mathrm{TL}}\ge 0$ is a tuning paramete. Here  $\Phi^{(t)}\in\mathbb{R}^{p_t\times p}$ is a matrix such that $ \bigl[\Phi^{(t)}\bigr]_{i,j\in\mathcal{R}(t)} = I_{p_t}$ with $I_{p_t}\in \mathbb{R}^{p_t\times p_t}$ denoting an identity matrix, and $\bigl[\Phi^{(t)}\bigr]_{i\in\mathcal{R}(t),\,j\in[\mathcal{R}(t)]^{c}} = \bigl(\Sigma_{\mathcal{R}(t)}\bigr)^{-1}\Sigma_{\mathcal{R}(t),[\mathcal{R}(t)]^{c}}$, where $[\mathcal{R}(t)]^c = [p]\setminus\mathcal{R}(t)$ denotes the set of features unobserved at the target site.\\

For a new independent test observation \((\tilde x_{\mathcal R(t)},\tilde y^{(t)})\) from the target site, we define the predictive risk of $\hat\omega_{\mathrm{TL}}^{(t)}$ as
\[
r^{\mathrm{pred}}\!\bigl(\hat\omega_{\mathrm{TL}}^{(t)}\bigr)
=
\mathbb E\!\left[
\bigl(\tilde y^{(t)}-\tilde x_{\mathcal R(t)}^\top \hat\omega_{\mathrm{TL}}^{(t)}\bigr)^2
\,\middle|\,
X_{\mathcal R(t)}^{(t)}
\right].
\]

Our goal is to derive an asymptotic upper bound for this predictive risk, as sharp as possible in the regime $n_t,p\to\infty$, together with the corresponding optimal choice of $\lambda_{\omega}$, under the following assumptions.\\

High-Dimensional Assumption:
The full-feature dimension $p$ and the observed-feature dimension $p_t$ go to infinity as $n_t$ goes to infinity, with $p/n_t\to \gamma \in (0,\infty)$ and $p_t/n_t\to \eta \in (0,\infty)$.\\

Random Regression Coefficients Assumption:
For each $m\in\{s,t\}$, the regression coefficient vector $\beta^{(m)}$ is random with $\mathbb{E}[\beta^{(m)}]=0$ and $\mathrm{Cov}[\beta^{(m)}]=(\alpha_m^2/p)I_p$, where $\alpha_m\neq 0$. In addition, $\mathrm{Cov}(\beta^{(t)},\beta^{(s)})=(\rho\alpha_t\alpha_s/p)I_p$, where $\rho \in(0,1]$.\\

Source Estimation Error Assumption:
The source estimator $\hat\beta^{(s)}$ satisfies $\lambda_{\max}\!\left(\mathbb E\bigl[(\hat\beta^{(s)}-\beta^{(s)})(\hat\beta^{(s)}-\beta^{(s)})^\top\bigr]\right)\le C^U/p$ uniformly as $(n_t,p)\to\infty$, for some constant $C^U>0$. Here $\lambda_{\max}(\cdot)$ denotes the largest eigenvalue.
\end{problembox}

We began with a zero-shot prompt providing the background context and problem setup, as shown in the box below.

\begin{promptbox}{Zero-shot prompt 3}
You are an expert in statistical methodology and mathematical reasoning. Your task is to produce a full formal proof for a problem in an ongoing research project.

\textbf{Background:} The background introduced in Context \ref{ctx:tl}

\textbf{Problem:} The problem stated in Problem \ref{prob:tl}.

Give a full formal proof for the problem.
You need to state assumptions and regularity conditions that you use.
Justify each step and give the complete proof in \LaTeX.

\end{promptbox}

Under this zero-shot prompt, the model decomposed the predictive risk and attempted to bound each term individually, a broadly reasonable high-level approach. However, the resulting bound was loose and uninformative. 
Two specific failures drove this outcome. First, for trace terms that admit an exact closed-form expression, the model instead applied a crude spectral bound, such as $\Sigma_{\mathcal R(t),\mathcal R(t)} \preceq C_\Sigma I_{p_t}$, where $C_\Sigma$ is the largest eigenvalue of $\Sigma_{\mathcal R(t),\mathcal R(t)}$. Second, for terms whose high-dimensional limits can be calculated exactly using tools from random matrix theory \citep{bai2010spectral}, the model did not attempt the calculation and instead simply assumed the limits as given. Overall, the result was a bound expressed in terms of unspecified quantities, making it difficult to understand how prediction risk depends on the key problem parameters, such as sample size, dimension, cross-site correlation, and feature mismatch.

We then explored whether the model could learn directly from existing literature that addresses similar problems using established domain principles. 
To this end, we supplemented the original prompt with targeted guidance drawn from two closely related papers. The first provided tools for exact trace calculations and deterministic equivalent approximations in high-dimensional settings \citep{dobriban2018high}. The second provided a risk decomposition and bounding strategy specifically developed for transfer learning estimators of this type \citep{gu2025robust}. Notably, we did not reproduce the arguments from these papers; we simply named them and indicated how they were relevant to the problem at hand.

\begin{promptbox}{Guidance prompt 2}
\textbf{Guidance:} 
Use the techniques listed below. (i) The limiting trace / deterministic equivalent calculations in Dobriban and Wager (2018), High-Dimensional Asymptotics of Prediction: Ridge Regression and Classification.
(ii) The predictive-risk decomposition and bounding strategy for the $L_2$ transfer learning estimator in Gu, Han, and Duan (2025), Robust Angle-Based Transfer Learning in High Dimensions.

Given the background, problem, and guidance above, give a full formal proof for the problem.
You need to state assumptions and regularity conditions that you use.
Justify each step and give the complete proof in \LaTeX.

\end{promptbox}

With this literature-based guidance, the model's performance improved substantially. Rather than applying loose spectral bounds throughout, it retained exact expressions for the terms that admit closed-form limits and applied bounding only where genuinely necessary, specifically, for the term involving source estimation error, where an exact expression is not available. The resulting asymptotic upper bound was sharper, more interpretable, and richer in statistical insight than the zero-shot results. 

This example highlights an important and practical lesson: in statistical proof development, the usefulness of an AI-generated result depends not only on its correctness but also on whether it leads to sharp and statistically meaningful conclusions. 
A loose but valid bound is not equivalent to a sharp and informative one: the former may be mathematically defensible but scientifically uninformative, failing to reveal how the quantity of interest depends on the key problem parameters. 
The model could bound terms but could not judge which admitted a closed-form limit and which could only be bounded, a distinction that requires domain expertise.
In our example, providing two targeted literature references was enough to supply that judgment, transforming a valid but uninformative derivation into one that is both rigorous and scientifically useful. 
Recognizing that a bound was loose rather than sharp, despite being mathematically valid, currently requires a human judgment of this kind: an expert who can quickly tell that a formally correct result is nonetheless scientifically unsatisfying, and who knows which literature would resolve the gap.

\subsection{Equivalence of privacy formalisms in differential privacy}
\label{positive example:data_privacy}

Our fourth example comes from an ongoing project on the exact relationships among different privacy formalisms in differential privacy (DP). Several standard languages exist for quantifying privacy guarantees, including R\'enyi-DP  \citep{bun2016concentrated}, $(\varepsilon,\delta)$-DP \citep{dwork2006our,dwork2006calibrating}, and $f$-DP \citep{Dong2022Gaussian, su20242020}. 
Each formalism arises from a different theoretical perspective and characterizes the privacy level through an exact curve, but these curves are indexed by different parameters and encode privacy from different viewpoints. 
Briefly, the R\'enyi-DP curve records the R\'enyi divergence as a function of the order $\alpha$, and thus defines privacy through a family of divergence comparisons between neighboring output distributions. The \((\varepsilon,\delta)\)-DP curve records, for each threshold $\varepsilon$, the smallest admissible failure probability $\delta$, so it defines the privacy through the tail behavior of the privacy loss. The \(f\)-DP curve is indexed by the type-I error level and gives the optimal type-II error for testing between two neighboring output distributions, defining privacy from a statistical distinguishability perspective.

The proof task is to investigate whether the full exact R\'enyi-DP curve determines the full exact \((\varepsilon,\delta)\)-DP curve and the full exact \(f\)-DP curve, and whether the converse implication holds. 
This question is natural and important because the same mechanism is often analyzed with different DP languages in the literature. 
For example, one may derive R\'enyi-DP to analyze composition, convert to \((\varepsilon,\delta)\)-DP for reporting guarantees, and use $f$-DP to interpret distinguishability \citep{abowd20222020}. A precise understanding of whether the corresponding full exact curves contain the same information is therefore necessary for translating privacy guarantees across formalisms without loss of information.
Specifically, we consider privacy additive-noise mechanisms with centered symmetric noise, of the form
\[
M(x)=A(x)+Z,
\]
where \(A\) is a deterministic algorithm and \(Z\) follows a centered symmetric noise distribution, such as a Gaussian or Laplace distribution \citep{dwork2006our,dwork2006calibrating,Mcsherry2010privacy}. 
Here $Z$ is the noise added to protect privacy.
The centered symmetry of $Z$ is not merely a technical convenience; it encodes a structural property of the mechanism that, as we show below, plays a decisive role in determining what privacy information can be recovered across formalisms.
The problem statement, context, and prompt provided to the model are shown in the boxes below.

\begin{problembox}{}{data_privacy}
Does the full exact R\'enyi-DP curve imply the full exact $(\varepsilon,\delta)$-DP curve and $f$-DP curve? Does the converse implication hold?
\end{problembox}

\begin{contextbox}{}{data_privacy}
Consider an additive-noise mechanism $M(x)=A(x)+Z$, where $A$ is deterministic and $Z\sim \nu$ is a centered symmetric noise distribution. Let $\mathcal{L}(X)$ denote the probability law of a random variable $X$. For neighboring datasets, define $P=\mathcal{L}(M(x))$ and $Q=\mathcal{L}(M(x'))$. We want to compare the information contained in the full exact R\'enyi-DP curve {$\{D_\alpha(P\|Q):\alpha>1\}$}, the full exact $(\varepsilon,\delta)$-DP curve {$\{\delta(\varepsilon): \varepsilon > 0\}$}, and the full exact $f$-DP curve {$\{f(\alpha): 1 \geq \alpha \geq 0\}$}.
\end{contextbox}

\begin{promptbox}{Zero-shot prompt 4}
``You are an expert in differential privacy and theoretical computer science. Your task is to produce a full formal proof for a problem in an ongoing research project.

\textbf{Problem}: The problem stated in Problem~\ref{prob:data_privacy}.

\textbf{Context}: The background provided in Context~\ref{ctx:data_privacy}.

Given the problem and context above, provide a rigorous mathematical answer. You need to state the assumptions and regularity conditions that you use, justify each step carefully, and give a complete proof in \LaTeX.''
\end{promptbox}

Under the zero-shot prompt, the model produced a partially correct answer. The forward direction was handled well: the model correctly argued that the full exact R\'enyi-DP curve contains enough information to recover the corresponding exact \((\varepsilon,\delta)\)-DP curve and the exact \(f\)-DP curve, successfully proving one side of the desired equivalence.
The failure occurred in the converse direction. 
Rather than exploiting the specific structure of the additive-noise mechanism with centered symmetric noise, the model treated the reverse implication as a generic question about arbitrary pairs of neighboring output distributions. 
Under that generic formulation, it concluded that the one-sided \((\varepsilon,\delta)\)-DP profile is strictly weaker and does not determine the other two curves, a conclusion that is reasonable in a general setting but misses the sharper result available here. Because the noise distribution is centered and symmetric, the one-sided \((\varepsilon,\delta)\)-DP profile in fact contains enough information to recover the full privacy description, making it equivalent to the full exact R\'enyi-DP and \(f\)-DP curves in this setting. 

To address this, we provided a short follow-up prompt explicitly directing the model back to the overlooked centered-symmetry assumption and asking it to reconsider the conclusion about one-sided versus two-sided privacy profiles, as shown in the box below.

\begin{promptbox}{Guidance prompt 3}
``Be careful that $Z\sim \nu$ is a centered symmetric noise distribution, so that reconsider your conclusion on the one-sided and two-sided $(\varepsilon,\delta)$-DP curve $\{\delta(\varepsilon)\}$.''
\end{promptbox}

With this targeted guidance, the model produced a correct and complete proof, concluding that
\begin{enumerate}
    \item For arbitrary \((P,Q)\), the usual one-sided profile \(\{\delta(\varepsilon):\varepsilon\ge 0\}\) is strictly weaker than the full exact R\'enyi-DP and \(f\)-DP curves; and 
    \item In the additive-noise setting with centered symmetric noise, the full exact R\'enyi-DP curve, the full exact \(f\)-DP curve, the full exact two-sided \((\varepsilon,\delta)\)-DP curve, and the usual one-sided exact profile \(\{\delta(\varepsilon):\varepsilon\ge 0\}\) are all equivalent.
\end{enumerate}

Unlike the other case studies, the limitation here was not a missing proof strategy or unfamiliar technique, but an incomplete use of the problem's key assumptions, leading the model to answer a more general question than the one being asked. In privacy protection, symmetry assumptions often encode nontrivial information about the relationship between neighboring output distributions, and this information can significantly affect what can be recovered from a given exact privacy curve \citep{vinterbo2022differential, balle2018privacy}. Overlooking such an assumption therefore does not merely weaken the proof; it changes the question being answered and the conclusion that can be drawn. This kind of incompleteness can be serious in research: an argument may be fully correct in a general setting and yet inadequate for the actual research question, because it misses a structural property that yields a stronger and more specific conclusion. Recognizing when a proof is answering the wrong question, and identifying which assumption restores the correct result, requires the kind of statistical judgment that current general-purpose LLMs do not reliably exercise on their own.
Yet this is also where human expertise can intervene efficiently: recognizing that the model had quietly answered a more general question than the one posed, a downstream diagnostic judgment rather than a strategic correction, took only a moment for a human expert. As this example demonstrates, a single short prompt highlighting the overlooked assumption was enough to redirect the model toward the correct and complete proof, turning an inadequate derivation into a rigorous one with minimal effort.

\section{When and how AI assistance fails}\label{notwork}

The four failure cases share a complementary pattern to the successful cases: current general-purpose LLMs break down most severely when the proof task is open-ended, the reasoning chain is long, or the problem requires simultaneous control of multiple technical components. In these settings, the execution–strategy gap widens: the model not only struggles to identify an appropriate proof strategy but also fails to sustain a coherent reasoning path long enough to reach a valid conclusion. Together, these cases reveal the current boundaries of general-purpose LLMs as proof assistants in statistical research, and point to the kinds of advances that would be needed to push those boundaries further.
They also sharpen the role of human expertise: as the model's own derivations grow less reliable, the burden of recognizing where and why an argument has gone wrong falls more heavily on the human collaborator.

\subsection{Asymptotic normality in high-dimensional inference}\label{negativeLDSC}

Our first failure case follows directly from the successful case in Section \ref{positveLDSC}. 
Having shown that AI models can handle a precisely scoped problem, computing the exact finite-sample variance of the oracle LDSC estimator, we now ask a more demanding question: can the model directly establish the asymptotic normality of the full LDSC estimator with given LD scores? This problem is substantially harder for at least two reasons. First, it requires a much longer reasoning chain, spanning multiple steps including variance calculation, denominator concentration, linearization of the ratio estimator, and a final normal approximation step. Second, it requires simultaneous control of multiple technical components that must be coordinated carefully across the full argument. The specific problem setup is described in the box below.

\begin{problembox}{}{probLDSCw}
	Suppose that we have a GWAS cohort with $\X\in \bbR^{n\times p}$ consisting samples of $p$ genetic variants and $\y\in\bbR^{n}$ consisting samples of the outcome. In addition, we have a reference panel with $\Z\in \bbR^{n_r\times p}$ consisting $n_r$ samples of the $p$ genetic variants. The GWAS cohort samples and reference panel samples are independent.
	Let $\bmSigma$ be the true covariance matrix of the $p$ genetic variants, and $\widehat{\bmSigma}$ be the corresponding sample covariance matrix.
	Define the estimated LD score $\hat\ell_{j}=\sum_{i=1}^{p}\widehat{\bmSigma}_{ji}^2$ and the estimated marginal covariance $\widehat{a}_j=n^{-1}\cdot \X_{j}^T\y$, where $\widehat{\bmSigma}_{ji}$ is the $(j,i)$ element of $\widehat{\bmSigma}$ and  $\X_{j}$ is the $j$-th column of $\X$ for $i=1, \dots, n$ and $j=1, \dots, p$.
	Let $\widehat{w}=(\widehat{a}_1^2,\dots, \widehat{a}_p^2)^\top$, $\hat \ell=(\hat\ell_{1},\dots,\hat\ell_{p})^\top$, 
	and
	\[
	\mu_{\hat\ell}=\frac{1}{p}\,\bm{1}_p^\top \hat\ell,\qquad
	\hat\mu_{w}=\frac{1}{p}\,\bm{1}_p^\top \hat w.
	\]
	Define the (centered) LDSC estimator
	\[
	\hat\sigma_\alpha^2
	= \frac{(\hat\ell-\mu_{\hat\ell}\bm{1}_p)^\top(\hat w-\hat\mu_{w}\bm{1}_p)}{(\hat\ell-\mu_{\hat\ell}\bm{1}_p)^\top(\hat\ell-\mu_{\hat\ell}\bm{1}_p)}.
	\]
	
	We assume that
	\[
	\y= \X\bmalpha+\bmeps
	\]
	where  $\bmalpha=(\alpha_1,\ldots, \alpha_{p})^{T}$ 
	are unknown fixed genetic effects and $\bmeps=(\epsilon_{1},\ldots,\epsilon_{n})^T$ represent mean-zero normal random error vectors. We also assume that all the genetic variants follow a multivariate normal distribution with a block-diagonal structured covariance matrix.
	
	Please show the asymptotic normality of $\hat\sigma_\alpha^2$ as $n,p\to\infty$.
\end{problembox}

We first applied a zero-shot prompt, then supplemented it with targeted guidance referencing two key papers, one for variance calculation and one for the asymptotic normality step, as shown in the boxes below.

\begin{promptbox}{Zero-shot prompt 5}
	You are an expert in mathematical statistics and genetics. 
	The following is a research-level problem. Please make the best effort to
	provide a rigorous and complete formal proof for the problem.
	
	\textbf{Problem:} The problem stated in Problem~\ref{prob:probLDSCw}
	
	\textbf{Emphasize:} (i) The genetic effects are fixed. Do NOT assume that they are random effects. 
	(ii) Please define notations carefully and use only  natural notations.
	(iii) Write the output in LaTeX language 
	using the standards of rigor and scholarship that
	prevail in the mathematical literature.
\end{promptbox}

\begin{promptbox}{Guidance prompt 4}
	Guidance: (i) For variance calculation, consider using Proposition S1 in the supplement of the paper ``Dicker, L. H. (2014). Variance estimation in high-dimensional linear models. Biometrika, 101(2), 269-284.'', which is attached. (ii) For asymptotic normality, consider using Theorem 2.2 in the paper ``Chatterjee, S. (2009). Fluctuations of eigenvalues and second order Poincaré inequalities. Probability Theory and Related Fields, 143(1), 1-40. (https://arxiv.org/abs/0705.1224)''.
\end{promptbox}

We tested all three models under the combined prompt. Despite the targeted guidance, all three produced polished-looking derivations that fell short of what would be required for use in statistical research. 
The Gemini 3.1 Pro output was the weakest. It provided only a high-level proof sketch, introduced additional scaling assumptions on the fixed effects that were not part of the original problem, and invoked Chatterjee's theorem \citep{chatterjee2009fluctuations} using heuristic derivative bounds rather than verified calculations. Several key claims, including the order of the variance, the concentration of the denominator of the LDSC estimator, and the derivative bounds needed to apply Chatterjee's theorem, were asserted with minimal justification. The argument had the right overall shape but remained far from a complete proof.

The GPT-5.4 Thinking and Claude Opus 4.6 outputs were substantially more polished, but still not ready for use in a research paper. Both models identified a plausible high-level pathway: linearize the ratio estimator, reduce the problem to a Gaussian fluctuation statement, and apply Chatterjee's theorem to conclude asymptotic normality. However, the main problem is that the most difficult parts of the argument were handled by introducing additional assumptions rather than being derived from the original setup. 
For example, in the GPT-5.4 Thinking output, the proof assumed the variance of the LD score estimators rather than deriving it from the model, a quantity that should follow from the setup, not be taken as given. The order of several derived bounds and estimates was also claimed without justification, leaving the final application of Chatterjee's theorem insufficiently supported. The Claude Opus 4.6 output went further but changed the target of the problem: rather than working directly with the original estimator, it introduced a pseudo-true centering quantity $\sigma_{\alpha,0}^2$ defined through the estimated LD scores and proved asymptotic normality around that modified target, shifting part of the original difficulty into new definitions and assumptions. Some crucial steps in the martingale-CLT argument were also only justified informally.

Across all three models, a consistent pattern emerged: once the reasoning path becomes long enough, the model tends to preserve the overall structure of a plausible proof by introducing convenient intermediate assumptions, auxiliary targets, or black-box asymptotic conditions at the points where the real derivation becomes most difficult. The proof appears complete on the surface, but the hardest steps have been quietly replaced by unjustified assumptions.
Combined with Section~\ref{positveLDSC}, this example suggests that current general-purpose LLMs are considerably more reliable on well-specified isolated problems than on full end-to-end asymptotic arguments. The execution–strategy gap widens as the reasoning chain lengthens: the model not only struggles to identify the right overall strategy but also loses the ability to sustain a rigorous argument across all the steps needed to reach a valid conclusion. 

\subsection{Strategy selection from weak cues in copula tail dependence}

Section~\ref{positive example:copula} showed that two short proof-strategy hints transformed a failed zero-shot attempt into a correct and rigorous proof. This raises a natural question: how much of that work are the hints actually doing? Could the model identify an appropriate strategy on its own if given weaker but still domain-relevant cues, without being told explicitly what to do?

To explore this, we returned to the same copula problem from Section~\ref{positive example:copula}, but this time replaced the explicit strategy hints with a short list of domain-relevant keywords: ``copula, tail dependence, extreme value theory, and tail dependence coefficient''. We asked the model to use these keywords, together with the problem statement and research context, to first propose its own proof strategy and then carry it out. The prompt is shown in the box below.

\begin{promptbox}{Keyword-guided prompt 1}
``You are an expert in statistical methodology and mathematical reasoning. Use the problem statement, the background context, and the given domain keywords to identify a proof strategy and then provide a full formal proof.

You must output  two labeled sections:

Strategy: A short paragraph describing the proof strategy.

Proof: A complete formal proof. State all assumptions and regularity conditions explicitly, justify each step, and prove the exact claim stated in the problem.''

\textbf{Problem:} The problem stated in Problem~\ref{prob:copula}.

\textbf{Context:} The background provided in Context~\ref{ctx:copula}.

\textbf{Keywords}: copula, tail dependence, extreme value theory, tail dependence coefficient.

Use the exact headers Strategy: and Proof:.''
\end{promptbox}

The model did propose a strategy below, but not the right one. Briefly, it suggested using the copula's coordinatewise monotonicity to sandwich the copula along the target path between two fixed multiplicative reparametrizations of $u$, then applying a squeeze argument to conclude the desired invariance. 
This approach does not invoke the triangle inequality or the uniform convergence arguments that the correct proof requires, and the resulting formal derivation introduced unverified assumptions and non-rigorous steps. In summary, the keywords pointed the model toward the right research area but provided no principled basis for selecting among the proof strategies available within that area. 

\begin{proofbox}{AI-generated strategy 1}
Use the copula's coordinatewise monotonicity to sandwich $C(\gamma u^\delta, \gamma^{-1} u^{2-\delta})$ between $C$ evaluated along the reference path $(\tilde{\gamma} (ru)^\delta, \tilde{\gamma}^{-1} (ru)^{2-\delta})$ at two fixed multiplicative reparametrizations $r=r_1$ and $r=r_2$ of $u$. 
Then prove that replacing $u$ by $ru$ does not change the asymptotic ratio $[\log C]/\log u$ when the limit for $\tilde{\gamma}$ exists. Apply a squeeze argument to conclude $\kappa_L(\delta,\gamma;C)=\kappa_L(\delta,\tilde{\gamma};C)$ for all $\gamma>0$.
\end{proofbox}

This experiment reveals a sharp distinction between two kinds of guidance. Domain keywords orient the model toward a relevant body of literature but leave the strategy selection problem unsolved: the model must still choose among many plausible approaches without the judgment to know which one fits. Explicit proof-strategy hints, by contrast, resolve that choice directly. This distinction reflects the execution-strategy gap in action: the model can follow a strategy once it is provided, but cannot select one reliably on its own, even when broad domain context is available.

\subsection{Multi-source transfer learning}\label{negative example:TLRisk}

Section \ref{positive example:TLRisk} showed that literature-based guidance could substantially improve the sharpness of an asymptotic bound in a single-source transfer learning problem. We now extend that problem to a more complex multi-source setting, asking whether the same guidance approach remains effective when the estimation procedure involves multiple stages and the reasoning chain grows correspondingly longer. The updated context and problem setup are described in the boxes below.

\begin{contextbox}{}{tl2}
We consider a transfer learning problem with a target site \(t\) and \(M\) source sites indexed by \(1,\dots,M\), where our primary interest is prediction at the target site. For each site \(m\in\{t,1,\dots,M\}\), the outcome \(y^{(m)}\in\mathbb{R}\) and the covariate vector \(x\in\mathbb{R}^p\) follow the linear model
\[
y^{(m)} = x^\top \beta^{(m)} + \varepsilon,
\]
where $\beta^{(m)}\in\mathbb{R}^p$ is a random coefficient vector, $x\sim N(0,\Sigma)$, and $\varepsilon\sim N(0,\sigma^2)$ is independent of $x$. The dimension $p$ increases to infinity as the sample size $n$ increases.
\end{contextbox}

\begin{problembox}{}{tl2}
Suppose that we only observe the subvector $x_{\mathcal R(m)}$, indexed by $\mathcal R(m)\subseteq [p]$ in each site $m\in\{t,1,\dots,M\}$. Let $X^{(m)}_{\mathcal R(m)}\in\mathbb{R}^{n_m\times p_m}$ and $Y^{(m)}\in\mathbb{R}^{n_m}$ denote the observed design matrix and response vector, respectively, where $p_m=|\mathcal R(m)|$ and $n_m$ is the sample size at the site $m$.\\

We propose an estimation procedure including 3 steps.
In the first step, we compute a local ridge estimator based on the observed data in each site $m\in\{1,\dots,M\}$:
\[
\widehat\omega^{(m)}_{\mathrm{Loc}}
:=
\arg\min_{\omega}
\left[
\frac{1}{n_m}\|Y^{(m)}-X^{(m)}_{\mathcal{R}(m)}\omega\|_2^2
+\lambda_{\omega}^{(m)}\|\omega\|_2^2
\right],
\]
where \(\lambda_{\omega}^{(m)}\ge 0\) is a tuning parameter.\\

In the second step, we define an aggregator that reconstructs a full-feature coefficient vector:
\[
\hat\beta^{(s)}
:=
\arg\min_{\beta}
\sum_{m\in[M]}
\bigl\|
\Phi^{(m)}\beta-\widehat\omega^{(m)}_{\mathrm{Loc}}
\bigr\|_{W_m}^2,
\]
where $W_m\succeq 0$ is positive semi-definite and $\|x\|_{W_m}=\langle x,W_mx\rangle$ for any vector $x$. Here $\Phi^{(m)}\in\mathbb R^{p_m\times p}$ is a matrix such that $[\Phi^{(m)}]_{i,j\in\mathcal R(m)}=I_{p_m}$ with $I_{p_m}\in\mathbb R^{p_m\times p_m}$ denoting an identity matrix, and $[\Phi^{(m)}]_{i\in\mathcal R(m),j\in[\mathcal R(m)]^c}=(\Sigma_{\mathcal R(m),\mathcal R(m)})^{-1}\Sigma_{\mathcal R(m),[\mathcal R(m)]^c}$, where $[\mathcal R(m)]^c=[p]\setminus\mathcal R(m)$ denotes the set of features unobserved at the site $m$.\\

In the third step, we let $\hat\omega^{(t)}_s=\Phi^{(t)}\hat\beta^{(s)}$ and perform transfer learning to obtain $\hat\omega_{\mathrm{TL}}^{(t)}$ [...]. \\

For a new independent test sample \((\tilde x_{\mathcal R(t)},\tilde y^{(t)})\) from the target site, we define a predictive risk for the transfer learning estimation above as
\[
r^{\mathrm{pred}}\!\bigl(\hat\omega_{\mathrm{TL}}^{(t)}\bigr)
=
\mathbb E\!\left[
\bigl(\tilde y^{(t)}-\tilde x_{\mathcal R(t)}^\top \hat\omega_{\mathrm{TL}}^{(t)}\bigr)^2
\,\middle|\,
X_{\mathcal R(t)}^{(t)}
\right].
\]
Our goal is to derive an asymptotic upper bound for the above predictive risk, as tight as possible as $n_t,p\to\infty$, with optimal choice of $\lambda_{\omega}^{(m)}$, $\lambda_{\mathrm{TL}}$ and $W_m$ under the high-dimensional and random-coefficient assumptions as below. Here the optimal $W_m$ refers to the aggregation weights that minimize the aggregation error $\mathbb{E}\left[(\hat\beta^{(s)}-\beta^{(t)})(\hat\beta^{(s)}-\beta^{(t)})^\top\right]$ in the Loewner sense.\\

High-Dimensional Assumption:
For $m\in\{t,1,\dots,M\}$, the full-feature dimension $p$ and the observed-feature dimension $p_m$ go to infinity as $n_m$ goes to infinity, with $p/n_m\to \gamma \in (0,\infty)$ and $p_m/n_m\to \eta_m \in (0,\infty)$.\\

Random Regression Coefficients Assumption:
For $m\in\{t,1,\dots,M\}$, the regression coefficient vector $\beta^{(m)}$ is random with $\mathbb{E}[\beta^{(m)}]=0$ and $\mathrm{Cov}[\beta^{(m)}]=(\alpha_m^2/p)I_p$, where $\alpha_m\neq0$. In addition, for any $m,m'\in\{t,1,\dots,M\}$, the coefficients $\beta^{(m)}$ and $\beta^{(m')}$ satisfy $\mathrm{Cov}(\beta^{(m)},\beta^{(m')})=(\rho_{m,m'}\alpha_m\alpha_{m'}/p)I_p$, where $\rho_{m,m'}\in[0,1]$. 
\end{problembox}

The $[...]$ in the prompt above represents the transfer learning procedure defined in Problem~\ref{prob:tl}. 
The extended problem introduces a third step: an aggregation stage that combines local ridge estimators from multiple source sites into a single full-feature coefficient vector before transfer learning is performed. 
The key challenge in this problem is to quantify how estimation error accumulates and propagates across each stage of the procedure \citep{li2022transfer,chang2024heterogeneous,tian2023transfer}. 
Therefore, a complete proof would proceed in three steps: (i) deriving the estimation error of the local ridge estimators at each source site; (ii) quantifying how this error propagates through the aggregator; and (iii) incorporating the resulting aggregation error into the final analysis of the transfer learning estimator at the target site. Each step builds on the previous one, and the final asymptotic upper bound for the predictive risk should be traceable back through all three stages to the original problem parameters.

To reflect this added complexity, we supplemented the guidance from Section~\ref{positive example:TLRisk} with a third literature reference specifically covering the error analysis of the aggregator and the optimal choice of weighting matrix. The guided prompt is shown in the box below. Here the Guidance references (ii) and (iii) were carried over from Section~\ref{positive example:TLRisk} to maintain continuity with the domain principles established there, while Guidance (i) is new to this example.

\begin{promptbox}{Guided prompt 5}
You are an expert in statistical methodology and mathematical reasoning. Your task is to produce a full formal proof for a problem in an ongoing research project.

\textbf{Background:} The background introduced in Context \ref{ctx:tl2}

\textbf{Problem:} The problem stated in Problem \ref{prob:tl2}.

\textbf{Guidance:} Use the techniques listed below. (i) The error analysis of the aggregator and the solution of optimal weighting metrics in Cheng, C., Cheng, G., \& Duchi, J. C. (2023). Collaboratively learning linear models with structured missing data. Advances in Neural Information Processing Systems, 36, 7529-7540. (ii) The limiting trace / deterministic equivalent calculations in Dobriban and Wager (2018), High-Dimensional Asymptotics of Prediction: Ridge Regression and Classification.
(iii) The predictive-risk decomposition and bounding strategy for the $L_2$ transfer learning estimator in Gu, Han, and Duan (2025), Robust Angle-Based Transfer Learning in High Dimensions.

Give a full formal proof for the problem.
You need to state assumptions and regularity conditions that you use.
Justify each step and give the complete proof in \LaTeX.

\end{promptbox}

The guided proof failed in a revealing way. Rather than deriving the estimation errors of the local ridge estimators and the aggregator from the model, the model assumed their convergence and defined the limits. Specifically, it assumed the convergence of the source estimation error \(\mathrm{Cov}(\widehat\omega_{\mathrm{Loc}}^{(m)}-\omega^{(m)})\to V_m\) with a deterministic matrix \(V_m\succeq0\). Similarly, it further assumed the error of aggregator, \(\mathrm{Cov}(\hat\omega^{(s)}-\omega^{(s)})\to C_A\) with a deterministic matrix \(C_A\succeq0\). Here \(V_m\) and \(C_A\) serve as unspecified covariance matrices and were never derived or bounded: they simply appeared as placeholders. As a result, the final asymptotic upper bound for the predictive risk was expressed in terms of these unspecified matrices, making it impossible to understand how the bound depends on the key problem parameters.

This example provides insights into the boundaries of literature-based guidance. In Section \ref{positive example:TLRisk}, the same guidance approach succeeded because the proof involved a manageable number of coordinated components. Here, the reasoning chain was longer and more interdependent, and the model could not track all the dependencies simultaneously. Even with targeted references pointing to the right tools, it resorted to replacing the hardest derivations with unverified assumptions. This suggests that the effectiveness of literature-based guidance may diminish as problem complexity grows, and that proof decomposition could be a more robust strategy for problems of this kind.

\subsection{Zero-concentrated differential privacy}
\label{sec:main-zcdp-gap}
Section~\ref{positive example:data_privacy} showed that the three standard privacy formalisms, R\'enyi-DP, \((\varepsilon,\delta)\)-DP, and \(f\)-DP, are equivalent under the centered-symmetry assumption when one retains the full exact privacy curve. We now ask a related but harder question: what happens when the full exact curve is replaced by a scalar summary? 
Specifically, we consider zero-concentrated differential privacy (zCDP) \citep{bun2016concentrated}, and a mechanism is said to satisfy \(\rho\)-zCDP if, for every order \(\alpha>1\), its R\'enyi divergence is bounded by
\[
D_\alpha(P\|Q)\leq \rho \alpha
\]
for every pair of neighboring output distributions \(P\) and \(Q\). 
Thus, zCDP can be viewed as a scalar summary of the full R\'enyi-DP curve: instead of looking at the entire function \(\alpha\mapsto D_\alpha(P\|Q)\), it keeps a single parameter \(\rho\) that upper bounds the curve by a line.
This scalar summary is convenient, as the composition rule is simple and additive,  but potentially lossy, since it replaces the full R\'enyi-DP curve with a single linear upper bound $ \rho \alpha$. The question is: how lossy is it, and under what conditions?
In particular, we ask when zCDP is nearly tight, as it is known to be for Gaussian mechanisms, and when it loses substantial information relative to the full exact  R\'enyi-DP curve, especially under non-Gaussian noise distributions and heterogeneous composition.   

Unlike the problem in Section~\ref{positive example:data_privacy}, this question is inherently more open-ended and has no pre-determined solution path. There is no standard proof strategy to retrieve, no directly applicable theorem to invoke, and no existing literature that resolves the question. The problem asks the model not to execute a known argument but to navigate an open solution space and identify a promising direction, a fundamentally different and more challenging task.

The problem statement, research context, and formal proof instruction are shown in the boxes below. We asked the model to derive lower bounds for a gap metric that explicitly quantifies the optimality of zCDP. Specifically, the bound should reveal how this optimality varies with the number of compositions, the allocation of privacy parameters, and the underlying noise distribution.

\begin{problembox}{}{data_privacy_more}
Consider a heterogeneous composition of mechanisms \[ M=(M_1,\dots,M_n), \] where each \(M_i\) uses the same noise family \(\nu\) (possibly non-Gaussian), and you may assume the noise is centered and symmetric whenever needed. Suppose each \(M_i\) satisfies \(\rho_i\)-zCDP, so the usual composition rule gives total privacy parameter \(\rho=\sum_{i=1}^n \rho_i\). The main question is: How can we rigorously quantify its optimality or sub-optimality as a function of the number of compositions \(n\), the allocation \((\rho_1,\dots,\rho_n)\), and the underlying noise distribution \(\nu\)?

More specifically, consider the following metric $$\Gamma_{n,\nu}(\alpha) := \rho \cdot \alpha - D_\alpha(P|Q), $$ 
or, depending on normalization, possibly 
$$\Gamma_{n,\nu}(\alpha) := \rho \cdot \alpha - \sum_{i=1}^n D_\alpha(P_i|Q_i)$$ 
that measures the gap between the zCDP accounting and the exact privacy level of \(M\). Your answer should derive lower bounds under suitable assumptions, explain how it depends on the number of compositions \(n\), the allocation \((\rho_1,\dots,\rho_n)\), and the underlying noise distribution \(\nu\)?

\end{problembox}

\begin{contextbox}{}{data_privacy_more}
A guiding principle is that zCDP is easy to compose but may lose information: in general, from \(\rho_i\)-zCDP one only gets total \(\rho=\sum_i \rho_i\), while for Gaussian mechanisms this accounting is known to be tight. In additive symmetric mechanisms, the exact privacy descriptions through the full R\'enyi DP curve, the full exact \((\varepsilon,\delta)\)-DP curve, and the \(f\)-DP curve are expected to be mathematically equivalent under suitable regularity assumptions. Thus the loss in zCDP accounting should be understood as the loss incurred by replacing the full exact privacy curve by a single scalar.
\end{contextbox}

\begin{promptbox}{Zero-shot prompt 6}
``You are an expert in differential privacy and theoretical computer science. Your task is to produce a full formal proof for a problem in an ongoing research project.

\textbf{Problem}: The problem stated in Problem~\ref{prob:data_privacy_more}.

\textbf{Context}: The background provided in Context~\ref{ctx:data_privacy_more}.

Given the problem and context above, provide a rigorous mathematical answer. You need to state the assumptions and regularity conditions that you use, justify each step carefully, and give a complete proof in \LaTeX.''
\end{promptbox}

The model's response did not address this open-ended problem. A meaningful answer is expected to reveal how the optimality of zCDP depends on the noise distribution \(\nu\) through interpretable quantities, such as the distance from Gaussianity, tail behavior, cumulants of the privacy loss, or curvature of the R\'enyi-DP curve, so that the bound becomes informative about when zCDP is nearly tight and when it loses substantial information.
The model instead introduced a ``local non-Gaussian envelope defect'' assumption with unspecified constants \(c_\nu(\alpha)\), \(C_\nu(\alpha)\), and \(r_0\), and derived a lower bound of the form \(\Gamma_{n,\nu}(\alpha)\ge c_\nu(\alpha)\sum_i\rho_i^2/2\). 
This displaced the core difficulty into undefined constants: since \(c_\nu(\alpha)\) and \(C_\nu(\alpha)\) were never computed from any interpretable privacy quantity, the bound has a formal shape but no scientific content. It does not reveal why zCDP is tight for Gaussian mechanisms and loose for non-Gaussian ones, the insight the problem was designed to extract.

The response also spent considerable effort on proving statements that are either textbook facts or not useful for this research question. For example, it provided a full proof that R\'enyi divergence tensorizes under independent composition, namely \(D_\alpha(P\|Q)=\sum_iD_\alpha(P_i\|Q_i)\), and later used basic tools such as Cauchy--Schwarz and Jensen's inequality to rewrite the bound in terms of \(\sum_i\rho_i^2\) or an effective number of compositions. They are well-known facts that do not advance the question of when zCDP is informative. The notation and writing style were also unnecessarily heavy. For example, the output introduced many auxiliary objects, such as \(\bar\rho_i\), \(g_\nu(\alpha;r)\), \(P_{\nu,r}\), \(Q_{\nu,r}\), \(H(w)\), \(n_{\mathrm{eff}}\), \(K_t(\lambda)\), and several envelope constants, without using them to illuminate the underlying mechanism. Such notation-heavy presentation makes the proof difficult to read and is not consistent with good scientific writing, where new notation should simplify the argument rather than obscure the main idea.

This example illustrates a failure case for open-ended research questions: the model produced something that resembles a proof, with assumptions, formal derivations, and derived bounds, but that does not engage with the actual research question. Faced with a large and open solution space, the model defaulted to local plausibility: making each individual step look reasonable without ensuring that the overall argument addresses what was asked. It suggests that open-ended research questions, where the solution path is not pre-determined and deep conceptual insight is required, currently may lie beyond the reliable reach of general-purpose LLMs.

\section{Toward AI-assisted statistical proof}

The case studies presented in this paper point to an actionable picture of where current general-purpose LLMs stand as proof assistants in statistical research. They are genuinely useful, capable of carrying out substantial and technically demanding derivations when the problem is precisely specified and the proof strategy is clear. But they are also systematically limited by the execution-strategy gap: the inability to identify, select, and adapt appropriate proof strategies on their own when facing a new problem and context. This gap is not merely a performance shortcoming; it reflects something deeper about what statistical proof requires that current general-purpose LLMs, trained broadly across many domains, are not yet equipped to provide. The gap, however, is only half the story. As Section~\ref{sec:overview} highlighted, the demands on human expertise persist on either side of it in the human-AI collaborative workflow: in formulating a good proof task, and in verifying whether what an AI model returns is valid, on-target, and sharp. At the level of individual research practice, they suggest concrete ways to structure AI-assisted proof workflows so that the model's strengths are leveraged and human judgment is applied where it matters most. At the level of the broader statistical community, they point to a coordinated agenda: building shared resources that make proof strategies more accessible, developing more capable AI systems that can reason across longer and more complex arguments and support faster verification, and training a new generation of statisticians to work effectively in human-AI collaboration. In this section, we discuss each of these directions.

\subsection{Practical lessons for individual statisticians}\label{roadmap_prompt}

Our case studies show that the difference between a useful and an unhelpful proof generated by a general-purpose LLM often comes down to how the human-AI interaction is structured. Four practical lessons emerge consistently across the examples.

The first is precision in problem specification. A well-structured prompt makes the target quantity, notation, modeling assumptions, asymptotic regime, and any nonstandard conventions explicit from the outset. When the setup is underspecified, the model tends to fill in the gaps with mathematically convenient but scientifically inappropriate defaults, as illustrated in Section~\ref{positveLDSC}, where an underspecified prompt led the model to adopt a random-effects framework inconsistent with the intended fixed-effects setup. Moreover, when the intended formulation departs from the default convention in the relevant literature, stating the correct assumption is often not enough, the prompt should also explicitly rule out the incorrect default. Negative instructions of this kind are an effective way to prevent the model from defaulting to a confident but wrong formulation.

The second lesson is the value of targeted guidance. When zero-shot prompting fails because the model cannot identify or retrieve the right proof strategy, providing concise strategy hints or key literature references can make a substantial difference, often transforming a non-rigorous or uninformative derivation into a correct and sharp one, as illustrated in Sections~\ref{positive example:copula} and~\ref{positive example:TLRisk}. The guidance does not need to be extensive: a few targeted sentences pointing to the right approach, or two or three literature references indicating the relevant tools, are often sufficient. What matters is that the guidance resolves the strategy selection problem that the model cannot solve on its own.

The third lesson is proof decomposition. When the proof task is long, open-ended, or requires simultaneous control of multiple technical components, asking the model to produce a complete end-to-end proof in a single pass is unlikely to succeed, as illustrated in Sections~\ref{negativeLDSC} and~\ref{negative example:TLRisk}.  A more effective approach is to decompose the proof into smaller and more tractable problems and work on each within a structured human-AI workflow. This decomposition strategy shifts the model's role from end-to-end proof generator to targeted proof assistant, a role it is considerably better equipped to play in human-AI collaboration.

The fourth lesson is that there is no magic prompt and that iterative engagement is needed. Even for the small and tractable problems illustrated in our successful case studies, a single prompt attempt was rarely sufficient on its own; the prompt blocks used to present our case studies, including zero-shot attempts and targeted guidance, are a simplification chosen for clarity of exposition, not a description of how the underlying work actually proceeded. Arriving at each guided result typically took more exploration than the final presentation shows. For more complex and open-ended problems, the need for iteration is even greater. The realistic workflow is therefore closer to an engaged research conversation than to a single prompt query: specify the question, ask the model to propose a proof route, inspect that route, verify the AI-generated output at each step, and repair what fails. Treating prompt design as a one-time act risks the failure mode our case studies observed: a plausible-looking derivation accepted without the iterative scrutiny that catches when it has gone wrong.

In summary, these four lessons suggest a practical workflow for AI-assisted statistical proof: specify the problem precisely, provide targeted guidance when the strategy is not obvious, decompose complex arguments into manageable problems, and treat the whole process as iterative rather than a single exchange. This workflow does not replace human statistical expertise. It redefines the human role, redirecting expertise away from routine execution and toward the judgments and decision-making that current AI cannot reliably make on its own.

\subsection{Building shared resources for proof strategies}\label{shared_resources}

The practical lessons in Section \ref{roadmap_prompt} all depend, to varying degrees, on the availability of the right proof strategy at the right moment. In the copula example of Section~\ref{positive example:copula}, two short strategy hints transformed a failed derivation into a correct proof. In the transfer learning example of Section~\ref{positive example:TLRisk}, two literature references were enough to shift the model from a loose and uninformative bound to a sharp and meaningful one. In these cases, the critical resource was not computational power or mathematical ability, it was access to the right domain-specific proof strategy, supplied by a human expert who knew where to look. 

This points to a broader opportunity for the statistics community. Statistical proof development has two distinctive components that set it apart from most mathematical proof: translating a real-world scientific question into a precise statistical framework with appropriate assumptions, and then identifying and adapting the right strategy from a repertoire of reusable domain-specific tools. Both components currently rely on knowledge that is rarely documented in a retrievable form. The modeling component covers decisions about which assumptions are scientifically appropriate, which quantities are meaningful, and which mathematical idealizations are justifiable. This knowledge is largely transmitted through domain-specific training and research experience, with little explicit guidance available for researchers working at the boundaries of their expertise. The strategy component covers knowing which proof technique applies to a given problem and how to adapt it to a new context. This knowledge is similarly scattered across textbooks, papers, and the tacit knowledge of experienced researchers, rarely organized in a way that makes it easy to find and reuse. Building shared, organized repositories that address both components could make this knowledge substantially more accessible across the community. Such repositories would document not only reusable proof strategies but also the modeling principles and assumption choices that characterize different research areas, indexed by proof goal, scientific context, and typical assumptions, with representative references for each entry.

Such repositories would serve multiple purposes simultaneously. For individual researchers, they would reduce the time spent searching for applicable proof techniques, particularly for researchers working at the boundaries of their expertise. For graduate students, they would provide a structured entry point into the proof strategies of a new field, complementing what is currently transmitted slowly and unevenly through textbooks and one-on-one advising. For AI systems, they could serve as a retrieval resource that addresses the strategy-selection bottleneck directly, providing the targeted guidance that our case studies show is often sufficient to unlock reliable AI performance. Recent community efforts such as \citet{Lu2025StatEval} and StatProver\footnote{\url{https://huggingface.co/datasets/StatAILab/StatProver}.} provide encouraging early examples of what such resources might look like, and point toward a more systematic and community-wide effort.

Building and maintaining such repositories will require collective effort from the statistics community, contributions from researchers across subfields, curation to ensure quality and consistency, and regular updates as new proof techniques emerge. We view this as an important and tractable community agenda item, one that would benefit both human and AI-assisted statistical proof development simultaneously.

\subsection{Agentic AI systems and formal verification}\label{agentic_AI_systems}

The workflow insights and shared repositories described in Sections~\ref{roadmap_prompt} and~\ref{shared_resources} can substantially improve the reliability of current general-purpose LLMs for well-scoped statistical proof tasks. 
But our failure cases, such as Sections~\ref{negativeLDSC} and~\ref{sec:main-zcdp-gap}, suggest that some limitations run deeper than what better prompting or richer resources can address. When the reasoning chain is long or when the problem is genuinely open-ended, current general-purpose LLMs struggle in ways that targeted guidance alone cannot resolve. These cases point to the need for more capable end-to-end AI architectures, such as agentic AI systems that can plan, maintain memory across steps, verify intermediate results, revise their own arguments, and draw on external tools and literature in a coordinated way.

Recent progress in agentic AI for mathematics provides an encouraging reference point. Systems such as Aletheia \citep{Feng2026Aletheia} and the AI co-mathematician \citep{Google2026CoMathematician} organize mathematical reasoning through iterative generation-verification-revision workflows, in which a generator proposes a candidate argument, a verifier checks for errors and logical gaps, and a reviser improves the argument based on the feedback. These systems also use web search and literature navigation to reduce fabricated citations and ground arguments in existing results. Applied to research-level mathematical problems, they have shown substantially stronger performance than single-pass LLM prompting on complex reasoning tasks \citep{Xia2026Neurosymbolic}. 
In addition, the use of formal proof assistants such as Lean\footnote{\url{https://lean-lang.org/}} can provide machine-checkable verification of mathematical arguments. Recent work has shown that LLMs can be trained to generate Lean-verifiable proofs for Olympiad-level problems, suggesting that formal verification could eventually serve as a rigorous and automated checking mechanism within agentic statistical proof workflows. This is one of the supports our case studies suggest are most needed: not a replacement for the rapid diagnostic judgment described in Section~\ref{sec:overview}, but a complementary check that can flag what even a careful human reviewer might miss in a long or notation-heavy derivation.
However, many of these advanced agentic AI systems and LLM-Lean integrations are not publicly available; they are developed and maintained by large research organizations with access to substantial computational resources, and are not yet accessible to everyday statistical researchers in the way that general-purpose LLMs are. Closing this accessibility gap is itself an important community challenge, and points to the value of open and community-driven development of agentic AI proof tools tailored to the needs of statistical research.

Notably, adapting these architectures to statistical proof development would require going beyond what current mathematical agentic AI systems provide. Statistical proofs are shaped by modeling assumptions tied to scientific application domains and often require careful domain-specific judgment about what constitutes a sharp and meaningful result. 
A statistical proof AI system would therefore need to integrate several capabilities that current systems lack: a mechanism for navigating and applying domain-specific modeling conventions, access to organized repositories of statistical proof strategies of the kind described in Section~\ref{shared_resources}, tools for verifying not just mathematical correctness but also the scientific meaningfulness and sharpness of conclusions, and an exposition-revision stage that simplifies notation, removes inessential definitions, and ensures that the final argument is readable and verifiable by a human statistician. 
Formal proof assistants such as Lean could play a valuable role in this architecture, providing a verification layer that catches logical gaps and unjustified steps of the kind observed in our failure cases. For higher-level assessment, such as whether a proof is asking and answering the right scientific question, human diagnostic judgment is unlikely to be easily replaced.

\subsection{Training future statisticians for AI-assisted research}

The findings of this study have direct implications for how the next generation of statisticians should be trained. Our case studies show that effective use of general-purpose LLMs in statistical proof development is not simply a matter of knowing how to operate the AI tools: it requires a set of intellectual skills that are distinct from, but deeply complementary to, the mathematical skills that current graduate training already emphasizes. The main educational goal is to train students to direct, constrain, and critically evaluate AI assistance, treating current general-purpose LLMs as proof assistants that are often useful for precisely formulated problems but unreliable for unsupervised end-to-end derivations. We suggest emphasizing five complementary skills that current curricula do not systematically address.

The first is precise problem formulation with AI interaction. Students should learn to identify important questions from a scientific context, formulate statistical problems with appropriate assumptions, and translate a research-level proof task into a prompt that conveys the problem precisely and unambiguously to the AI model. This skill is closely related to the broader statistical skill of problem formulation, but applied in a new context where the consequences of underspecification are immediate and sometimes difficult to detect. For example, the LDSC example in Section~\ref{positveLDSC} illustrates both the importance of this skill and the subtle pitfalls that arise when it is not exercised carefully. 
As Section~\ref{sec:overview} noted, this is also a skill whose importance will likely increase: as AI handles more of the execution, what distinguishes researchers is increasingly their ability to formulate the right question in the first place. From a technical perspective, learning structured prompting techniques can also help students formulate research problems more precisely with AI.
From a technical perspective, actively keeping up with fast-evolving structured prompting techniques\footnote{\url{https://github.com/mattpocock/skills}} can also sharpen students' ability to formulate research problems with AI.
For example, rather than relying solely on the student to write a complete and precise prompt, a ``grill me" routine asks the AI model itself to interrogate the student about the theorem statement, the assumptions, the target quantity, and the intended proof strategy, until the problem is well specified before any derivation begins.

The second is proof decomposition. Students should learn to identify which components of a proof are well-scoped enough for reliable AI assistance and which require human judgment, and to decompose complex arguments into tractable tasks accordingly. This skill is not only useful for AI-assisted proof development, but it is also a valuable approach in its own right, encouraging students to think carefully about the logical structure of a proof before attempting to execute it.

The third is critical evaluation of AI-generated proofs. Students should learn to read AI-generated derivations with the same critical eye they would apply to a proof from a textbook or a paper, checking for hidden assumptions, unjustified appeals to standard results, changes in the estimand, and notation that obscures rather than clarifies the argument. Our failure cases show that general-purpose LLMs can produce derivations that appear complete and rigorous on the surface but contain errors that are easy to miss and hard to detect without domain expertise. Actively learning structured prompting techniques can support this skill in practice. For example, a ``zoom out" routine can ask the model to explain the proof's overall architecture, the role of each assumption, and where the main difficulty lies, before technical details are generated, which helps a student check the shape of an argument before getting lost in its details. An ``assumption audit" routine can ask the model to list every assumption used in the proof and classify each one as stated in the original problem, derived during the argument, newly introduced, or unnecessary, which makes hidden or unjustified assumptions more visible and checkable. Detecting these errors requires both mathematical maturity and familiarity with the specific failure modes that current general-purpose LLMs tend to exhibit.

The fourth is strategic use of guidance. Students should learn to recognize when a zero-shot prompt is unlikely to succeed. This is often the case when the problem requires strategy selection, careful control of modeling assumptions, or a long reasoning chain. In such situations, students should learn to provide targeted guidance that resolves the current bottleneck efficiently. This includes knowing how to identify relevant literature references, how to formulate concise proof-strategy hints, and how to calibrate the level of guidance needed for a given task. The copula and transfer learning examples in Sections~\ref{positive example:copula} and~\ref{positive example:TLRisk} illustrate how modest but well-targeted guidance can transform an unreliable AI output into a useful and rigorous one.

The fifth is rapid diagnostic judgment. As Section~\ref{roadmap_prompt} discussed, what separates a productive AI-assisted session from an unproductive one is often not whether an error eventually gets caught, but how quickly it is caught and corrected within the interactive workflow. A student who can tell within moments that a derivation has changed its target, smuggled in an unjustified assumption, or produced a result that is technically valid but scientifically weak can redirect the model immediately and keep a multi-round conversation moving toward a correct proof. A student who lacks this fluency may accept a flawed intermediate step, build further reasoning on top of it, and only discover the error much later, at far greater cost. This is, at its core, a form of human decision making rather than purely a technical skill: the student must decide whether to trust an output, redirect it, or dig deeper before proceeding, decisions that AI cannot make on the student's behalf.

Put together, these five skills outline what a well-structured AI-assisted proof session might look like for future statisticians.
It begins with formulating the statistical question, sharpened where useful by a grill-me interrogation that forces precision before any derivation starts. The student then selects or proposes a proof strategy, supplying targeted guidance where zero-shot prompting is unlikely to succeed, and decomposing the argument into lemma-level problems where the reasoning chain is long or the components are numerous. As the model generates each piece of the proof, a zoom-out check confirms that the overall architecture is sound before the details are scrutinized, and an assumption audit surfaces anything stated, derived, added, or left unjustified along the way. Throughout, rapid diagnostic judgment operates continuously, catching a wrong target or a quietly weakened result early enough to redirect the conversation. The exposition is then refined once the argument itself is sound, simplifying notation and removing inessential definitions so that the final proof is as readable to a human reader as it is correct. None of these stages is mechanical on its own; each depends on the kind of judgment that AI assistance relocates to the human rather than removes.

These skills can be integrated into graduate training through dedicated courses on AI-assisted statistical proof. Such courses would be structured around hands-on exercises in which students practice the five skills described above: formulating precise proof tasks, decomposing complex arguments, evaluating AI-generated outputs critically, providing targeted guidance, and developing rapid diagnostic judgment. The goal is not to replace traditional proof training but to complement it, preparing future statisticians to work effectively at the interface of statistical expertise and AI assistance that will increasingly characterize statistical research in the years ahead. Since courses of this kind are new to most statistics departments, developing them will itself benefit from community-wide collaboration: sharing course materials, exercises, and assessment frameworks across departments, and revisiting them regularly as the capabilities of AI tools continue to evolve rapidly.

\section{Limitations and future work}
The practical insights of this paper should be interpreted in light of several limitations. 
The first limitation is scope. Our findings are based on eight proof problems drawn from four areas of statistical research. While these problems were chosen to reflect the range of tasks that arise in day-to-day statistical research, and while the qualitative conclusions were consistent across three different models, eight problems cannot capture the full diversity of proof tasks in statistics. The execution-strategy gap we identify may manifest differently, or may be less pronounced, in other fields, proof types, or levels of technical difficulty. Future work can extend this investigation to a broader range of problems, including problems from areas such as Bayesian statistics, causal inference, and survival analysis, where proof strategies and modeling conventions differ substantially from those examined here.

A second limitation is the rapid pace of AI development. The findings reported here reflect the capabilities of GPT-5.4 Thinking, Gemini 3.1 Pro, and Claude Opus 4.6 at the time of writing. Model capabilities are changing rapidly, and some of the limitations we observe may be partially addressed by future models. We view this study as a snapshot of the current state rather than a permanent characterization, and we plan to update our findings as new models and tools become available. This applies with particular force to our claim that the importance of problem formulation and rapid diagnostic judgment will grow as execution improves: this is a prediction about a dynamic, not a direct empirical finding from our case studies, and it deserves the same scrutiny and updating as any other forecast about a fast-moving technology. 
The paired case study design introduced here, with matched success and failure cases drawn from the same research context, provides a reusable framework for ongoing evaluation as the landscape evolves. 

Another limitation is that we focus exclusively on general-purpose LLMs accessible to everyday researchers. Domain-specific fine-tuned models and agentic AI architectures may perform substantially better on statistical proof tasks. As discussed in Section~\ref{agentic_AI_systems}, some early evidence from mathematical agentic AI suggests meaningful improvements on complex reasoning tasks. Evaluating these more capable but currently less accessible systems on statistical proof tasks is an important direction for future work, particularly after these systems become more publicly available.

Despite these limitations, we believe the findings reported here provide a useful and actionable starting point for the statistics community. The execution-strategy gap is a real and systematic phenomenon that shapes the reliability of current general-purpose LLMs on statistical proof tasks, and the practical lessons and community agenda outlined in this paper provide concrete directions for addressing it, both through better use of existing AI tools and through the development of more capable ones. More broadly, statistical proof serves as a window into what human-AI collaboration in expert cognitive work actually demands: structured, verifiable, and tied to real-world scientific questions, it makes visible patterns that are likely to recur across domains where experts delegate technical execution to AI while remaining responsible for its direction and outcome.

\section*{Acknowledgments}
The research has been partially supported by NSF Grant DMS 2210860 and start-up funds from Purdue Statistics Department and Department of Statistics and Data Science at the University of Pennsylvania. 
This research was supported in part by Analytics at Wharton, Wharton AI \& Analytics Initiative, Wharton Dean’s Research Fund, Perelman School of Medicine CCEB Innovation Center Grant, and the University Research Foundation Grant. This research was also supported by the National Institute of Mental Health under Award Number R01MH136055, and National Institute on Aging under Award Numbers RF1AG082938,  R01AG085581, and R03AG102039. The content is solely the responsibility of the authors and does not necessarily represent the official views of the National Institutes of Health. 

\section*{Author contributions}
B.Z. and F.X. designed the study. F.X., X.S., H.T, and B.S., performed the case study analysis and generated most of the results.
M.M and C.W. provided input on the research question, prompt design, and interpretation of the findings.
X.S., H.T, B.S., F.X., and B.Z. wrote the manuscript with feedback from all authors.

\section*{Supplementary materials}
Supplementary materials are available at \url{https://github.com/xiaojing777/Human-AI-Statistical-Proof-Supplement}.

\section*{Competing interests}
The authors declare no competing interests.

\bibliographystyle{plainnat}
\bibliography{References}

\end{document}